\documentclass[11pt]{article}
\usepackage{moriond,epsfig}

\def\Journal#1#2#3#4{{#1} {\bf #2}, #3 (#4)}

\def\NCA{\em Nuovo Cimento}

\def\PLB{{\em Phys. Lett.}  B}
\def\PRL{\em Phys. Rev. Lett.}
\def\PRD{{\em Phys. Rev.} D}

\def\be{\begin{equation}}
\def\ee{\end{equation}}
\def\bea{\begin{eqnarray}}
\def\eea{\end{eqnarray}}

\begin{document}
\vspace*{4cm}
\title{THE CRYOGENIC DARK MATTER SEARCH EXPERIMENT, RESULTS FROM THE 2004 CAMPAIGN AND STATUS OF THE CURRENT UPDATE}

\author{ S. LECLERCQ }

\address{University of Florida, Department of Physics, P.O. Box 118440,\\
Gainesville, FL, 32611-8440, USA}

\maketitle\abstracts{
The CDMS II experiment uses Z-dependent Ionization Phonon (ZIP) detectors made of Germanium and Silicon to identify nuclear recoils from Weakly Interacting Massive Particles (WIMPs) with near complete event-by-event rejection of various radioactive backgrounds.
In 2004 CDMS II operated 6 Ge ZIPs and 6 Si ZIPs. The 74.5 live days of operation gave after cuts 34 kg{$\times$}d exposure for the Ge ZIPs and 15 kg{$\times$}d exposure for Si ZIPs. All criteria for identifying a signal from nuclear recoil due to WIMPs were developed blind with respect to the WIMP search data. The new 90\% C.L. upper limit on the spin-independent WIMP-nucleon cross section is 1.6 $\times$ 10$^{-43}$ cm$^{2}$ from Si, for a WIMP mass of 60 GeV/c$^{2}$.
The experiment has recently upgraded to 19 Ge ZIPs totaling 4.8 kg, and 11 Si ZIPs totaling 1.9 kg. The goal is to increase sensitivity with running in 2006 and 2007 by one order of magnitude.}

\section{Cold dark matter particles and their direct detection}

A great variety of observations suggest the existence of nonbaryonic dark matter clustered in galactic halos. The cosmic microwave background anisotropies show evidence that it represents one quarter of the energy density of the universe. A viable dark matter candidate is a weakly interacting massive particle (WIMP) that would have decoupled from the primordial plasma when it was non-relativistic. A natural WIMP candidate is offered in the minimal supersymmetric standard model (MSSM) by the lightest supersymmetric particle (LSP or neutralino), expected to be stable if the R-parity is conserved~\cite{r118prl}. Another WIMP candidate comes from flat universal extra dimension (UED) models~\cite{acd}, where momentum conservation implies the existence of a stable lightest Kaluza-Klein particle (LKP). A typical WIMP has a scattering cross section with an atomic nucleus characteristic of the weak interaction and an expected mass of {$\approx$}10-1000 of GeV/c$^{2}$. The nucleon coupling in the extreme nonrelativistic limit is characterized by two terms: spin-independent (e.g. scalar) and spin-dependent (e.g. axial vector)~\cite{sdprd}. Nucleon contributions interfere constructively to enhance the WIMP-nucleus elastic cross section so that scalar couplings dominate direct-detection event rates in most models. In contrast, the axial couplings of nucleons with opposing spins interfere destructively, leaving WIMP scattering amplitudes determined roughly by the unpaired nucleons (if any) in the target nucleus. Spin-dependent interactions may dominate direct-detection event rates in regions of parameter space where the scalar coupling is strongly suppressed. Their amplitudes are also more robust against fine cancellations~\cite{sdprd}.\\
Rotation curve measurements suggest that in our galaxy WIMPs would have a mean velocity of {$\approx$}230 km/s and a local density of 0.3 GeV/cm$^{3}$. WIMP-nucleon elastic scattering would then occur in a 1-100 keV recoil energy range with a rate of less than 1 event keV/kg/day. In a direct detection experiment shielding has to be used to minimize backgrounds produced outside the apparatus. It is mandatory to discriminate between electron recoil events due to residual contaminants and nuclear recoil events~\cite{r118prl}$^{,}$~\cite{bgprd}$^{,}$~\cite{siprl}due to WIMPs and neutrons. Cosmic ray muons interacting with atmosphere and ground produce a neutron flux dominant at the Earth surface.

\section{The CDMS experiment}

The CDMS collaboration is operating the CDMS-II experiment in the Soudan Underground Laboratory (Minnesota, U.S.A.). The 780 m (2090 meters water equivalent) of rock overburden reduces the surface muon flux by a factor of 5{$\times$}10$^{4}$.\\
At the experiment's core, Z(depth)-sensitive ionization and phonon detectors (ZIPs) measure the ionization and athermal phonon signals caused by recoiling particles in Ge and Si crystals~\cite{bgprd}. Unlike neutrons, WIMP interactions would occur more often in Ge than in Si. The cold volume housing the detector towers, the attached $^{3}$He-$^{4}$He dilution refrigerator, and surrounding shielding are housed in a radio frequency (RF) clean room. Pumps, cryogenic supplies and control, and most of the electronics are situated outside of the RF room. The detector volume is the innermost of six nested copper cans that together make up the CDMS cryostat or "icebox". A set of concentric pipes allow to thermally couple the cans to the dilution refrigerator, and another set of copper pipes contains striplines connecting detectors to the room-temperature electronics. The icebox is surrounded by a 2-mm-thick mu-metal shield. Each tower has six detectors, and four temperature stages from 30 mK to 4 K. For each detector an electronics card, mounted on top of the tower, contains two field effect transistors (FETs) used for the readout of the ionization channels and self-heat to 130 K for optimal noise performance. A separate card at 600 mK contains four arrays of superconducting quantum interference devices (SQUIDs) required for the readout of the phonon channels. The cold hardware is constructed from radioactively-screened low background materials~\cite{bgprd}.\\
Concentrically arranged around the icebox, several layers are used to shield detectors from high-energy neutrons produced by close muon interactions, and from alphas, betas, gammas and neutrons from natural radioactivity. Outermost is an active muon veto made of forty 5-cm-thick plastic scintillator panels, connected to 2-inch photomultipliers able to distinguish between muons and ambient photons. The measured detection efficiency of the veto system is 99.4 $\pm$ 0.2\% for stopped muons and 99.98 $\pm$ 0.02\% for through-going muons. On average, one muon per minute is incident on the veto, but ambient gammas produce a veto rate of {$\approx$}600 Hz. Beneath the veto, a 40-cm-thick cylindrical polyethylene layer moderates low-energy neutrons from radioactive decays. It envelopes a 22.5-cm-thick lead shield, of which the inner 4.5 cm thickness consists of ancient lead. After 10 more cm of polyethylene, the icebox provides an average shielding thickness of about 3 cm of copper surrounding the detectors. Beginning on November  2003 the air volume between the icebox and the mu-metal shield has been continuously purged with "old air" to minimize activity from $^{222}$Rn and its associated daughters.

\section{ZIP detectors}

Each ZIP detector is a cylindrical high-purity 250 g Ge or 100 g Si crystal, 1 cm thick and 7.6 cm in diameter. Within a tower, the six ZIP detectors are stacked 2 mm apart with no intervening material. A particle scattering in a ZIP detector deposits energy into the crystal through charge excitations (electron-hole pairs) and lattice vibrations (phonons). Depending on the material and the type of recoil, 6\% to 33\% of the recoil energy is first converted into ionization before subsequent conversion to phonons. On average, one electron-hole pair is produced for every 3.0 eV (3.8 eV) of energy deposited by an electron recoil in Ge (Si). The ionization energy is defined as the recoil energy inferred from the detected number of charge pairs by assuming an electron recoil event with 100\% collection efficiency. The dimensionless ionization yield parameter, $y$, is the ratio of ionization energy to true recoil energy. This definition gives unity for electron-recoil events with complete charge-collection. The ionization yield for nuclear-recoil events is typically $y \approx 0.3$ in Ge and $y \approx 0.25$ in Si.\\
On each ZIP, the ionization sensors are composed by a disk-shaped inner fiducial electrode covering 85\% of the ionization side, and an annular outer guard ring used to reject events near the detector edges (where ionization and phonon responses are worse and background rates higher). The electric field created by the electrodes has to be set at less then few volt/cm to limit the Neganov-Trofimov-Luke (NTL) phonons~\cite{bgprd} produced by the electrons and holes motion. The recombination of the drifted charges in the electrodes releases all of the recoil energy from the electron system into the phonon system. There are two cases for which the ionization is underestimated. In the first case, impurity sites left with a net charge trap the drifting electrons or holes. We routinely use flashing LEDs to neutralize these sites. In the second case, the low electric field and self-screening from the initial electron-hole cloud enable some of the electrons or holes to drift into the ÒincorrectÓ electrode when events occur close to it. A thin layer ({$\approx$}40 nm) of doped amorphous silicon between each electrode and the detector surface reduced this effect, but events occurring within 10 $\mu$m depth still have a deficient charge collection.\\
Photolithographically patterned on the other side of the detector, a total of 4144 quasiparticle-assisted electrothermal feedback transition-edge sensors (QETs)~\cite{bgprd} form the phonon sensors. The QETs are divided into four independent channels. Each QET consists of a 1 $\mu$m-wide strip of tungsten (35 nm thick) connected to eight superconducting aluminum collection fins (300 nm thick), each roughly 380 $\mu$m $\times$ 55 $\mu$m. The tungsten strips form the transition-edge sensors (TESs), which are voltage biased, with the current through them monitored. The tungsten is maintained stably (T $\leq$ 50 mK) within its superconducting-to-normal transition (Tc {$\approx$}80 mK) by electrothermal feedback. Most phonons in the crystal that reach the aluminum fins scatter into them, creating diffusing quasiparticle excitations that later enter the tungsten TES, increasing the temperature of conduction electrons. The electrothermal feedback guarantees that the power deposited into the TES is exactly compensated for by a reduction in Joule heating. The energy in the phonon system includes not only the full recoil energy, but also energy from NTL phonons which contribute up to 50\% of the total. One of the advantages of measuring the athermal phonon signal is that it provides sensitivity to phonon physics that is dependent on the nature of the interaction and the event location. An interaction in the crystal produces high-frequency phonons which propagate quasidiffusively, combining elastic scattering and anharmonic decay. The number of phonons and their frequency decrease. Below 1 Thz they become ballistic: their mean free path becomes comparable to the detector size and they reach the speed of sound (5 mm/$\mu$s for Ge and 2.5 mm/$\mu$s for Si), 3 times faster than the high frequency phonons. The difference in propagation speed of the two phonon populations leads to a faster phonon leading edge for electron recoils when compared to nuclear recoils because of the larger fraction of NTL phonons which are all ballistic~\cite{bgprd}. When electrons and holes relax to the Fermi surface in the metal electrodes, most of the released energy is rapidly down-converted from high-frequency phonons to a third population of ballistic phonons. This process is all the more important that events are close to the detector surface and it produces ballistic phonons much more rapidly than the down-conversion from quasidiffuse propagation. Therefore the analysis of the phonon pulse shape makes it possible to reject surface events.\\
The CDMS detectors are made of natural Ge or Si, both composed predominantly of spinless isotopes,  sensitive to spin-independent interactions. However, each contains one significant isotope with nonzero nuclear spin: $^{73}$Ge (spin-9/2) makes up 7.73\% of natural Ge, while $^{29}$Si (spin-1/2) makes up 4.68\% of natural Si. Each isotope contains a single unpaired neutron, making CDMS sensitive to spin-dependent interactions~\cite{sdprd} with neutrons, but also with protons at a lesser extent.

\section{Calibrations, background and simulations}

A $^{133}$Ba source inserted above the icebox served to characterize the detector response to electron recoils. This isotope offers several distinct lines (276, 303, 356, and 384 keV) sufficiently energetic to allow the photons to penetrate the copper cans. The lines were very clear in all Ge detectors, including the 10.4 keV line from cosmogenically produced Ga. The comparison of the data to Monte Carlo simulations allowed to perform an accurate energy calibration. In Si detectors Compton scattering dominates at Ba line energies, the calibration is then achieved by matching the spectral shapes of the data and the simulation. We used a $^{252}$Cf source to characterize the detector response to nuclear recoils. The level of agreement between data and simulations suggests that the phonon measurement is independent of the recoil type.\\
Neutrons, gammas, betas, and alphas comprise the background of the experiment. Neutrons produce nuclear recoils and cannot be rejected on an event-by-event basis unless they scatter in more than one detector. We simulated the expected neutron background at Soudan~\cite{bgprd}, from U/Th nuclear decay chains and from muons, and found that it is insignificant for all CDMS runs. Gammas and betas can be misidentified as nuclear recoils when they scatter close to the surface of a detector. The majority of the gamma background interacts however in the bulk. The comparison of the WIMP-search electron recoils to simulation spectra allows us to identify the sources. A quarter of it seems to be due to remaining U/Th/K contaminants in the copper cans, and the rest is mostly due to Rn decay chain outside the purged volume. Betas arising from contamination on the detector surface are the most difficult background to characterize. There are a number of possible beta emitters such as $^{40}$K, $^{14}$C, and $^{210}$Pb. The combination of depth distribution simulations for several sources with a  model of the detector response (based on $^{109}$Cd data and simulations) shows that betas from surface contamination are the dominant electromagnetic background. Surface contaminants can also emit alphas, easily identifiable thanks to their large recoil energies, which is useful to estimate the amount of contamination that may produce other backgrounds. In particular  alphas from $^{210}$Po are a good tracer of the $^{210}$Pb decay chain.

\section{Data analysis}

For the analysis of the ionization waveforms, we constructed templates, both for the primary pulse and for crosstalk between electrodes, by averaging a number of ionization pulses from $^{133}$Ba calibrations, in the 10-100 keV range. We used two algorithms to estimate the amplitude and time offset of an ionization pulse. In the optimal-filter algorithm~\cite{bgprd}, the convolution of a filter (constructed in the frequency domain from the template and the noise spectrum) with an ionization pulse reaches a maximum which gives an estimate of the pulse height. In the second algorithm, a time-domain $\chi^{2}$ minimization of the template for each pulse estimates the amplitude. The time-domain algorithm is used for events saturating the digitizers ($\geq$ 1 MeV).\\
For the analysis of the phonon waveforms, we used a double exponential template with risetime {$\approx$}30 $\mu$s (15 $\mu$s) and falltime {$\approx$}300 $\mu$s (150 $\mu$s ) for Ge (for Si). For each phonon channel, we determine the energy using two different algorithms for different energy regimes. At relatively low energies ($<$ 100 keV) we use the optimal filtering technique, and at higher energies, where the phonon sensors saturates, we integrate  the pulse after subtraction of the average prepulse baseline. To construct relevant quantities for the position reconstruction, we estimate the times at which each pulse reaches 10\%, 20\%, and 40\% of the peak along the rising edge. We define the phonon "delay" time as the difference between the 20\% times of the phonon and ionization pulses. The phonon "risetime" is the difference between the 10\% and 40\% phonon times. The "peak" sensor is the channel that has the most energy. The relative phonon delays in the peak sensor and its two neighbors give a nonlinear mapping of the event position in the sensor plane. A second nonlinear measure of the position is determined from the relative partition of energy among the four phonon channels. The phonon timing and partition are not single-valued, but their combination allows an accurate parametrization of an event according to its physical position. The phonon pulse shapes depend on the event location within the detector for two reasons. First, variations in the TESs transition temperatures induce variations in their response. Second, the arrival times of phonons at the QETs depend on the position. These effects lead to variations in the timing parameters which would prevent the use of a single cut to discriminate surface events from bulk events. To remedy this, we apply position-dependent corrections making each parameter independent of the radial position. We correct for a nonlinearity of the phonon energy response, normalize the peak delay and peak risetime, and use a lookup table (average of the timing parameters of groups of 80 neighbors in a set of 12000 calibration events) to remove the position dependence of the energy-corrected quantities.\\
During the construction of the cuts of our analysis, the WIMP-search data was blinded to avoid human bias. We defined 10 main cuts. (1) The data-quality cut removes all non-optimal data (non-operational channels, noise bursts, etc.). (2) The phonon pretrigger cut rejects events with noisy pretrigger part of a phonon trace. (3) The ionization $\chi^{2}$ cut of the optimal-filter fit rejects events with anomalous ionization pulse shapes which can be from pile-up or noise glitches. (4) The fiducial-volume cut rejects events in the outer ionization electrode. (5) The electron-recoil bands and nuclear-recoil bands cuts are defined as the {$\pm$}2$\sigma$ widths of Gaussian fits to the distributions of ionization yield for both electron-recoil ($^{133}$Ba) and nuclear-recoil ($^{252}$Cf) events in several recoil-energy bins. (6) The ionization threshold cut selects events that have measurable pulses in the charge channels. (7) The muon-veto cut rejects events for which the global trigger occurred in the 50 $\mu$s after veto activity (remove $>$99.4\% of events caused by muons). (8) The single scatter cut selects events in which only one detector had a phonon signal larger than 6$\sigma$ of the energy distribution for random noise. (9) The timing cut rejects surface events. Limits in the peak delay and peak risetime plane, for several energy bins, were defined so that all events in the 4$\sigma$ nuclear-recoil band of Ba calibrations were rejected. Because of the QET $T_c$ distribution, lookup tables, and variations of the ratio of surface events during the calibration, the discrimination power of the timing cut varies from detector to detector. The leakage of the timing cut is estimated by testing it on a set of Ba calibration data not used in its definition. In 2005 we completed five distinct timing analysis to improve the rejection of surface events~\cite{siprl}. One uses the sum of delay and risetime. Two other evaluate the $\chi^{2}$ of the fit for surface versus bulk events, using the time delay, risetime and energy partition variables ; one of them is energy-dependent. Another one combines delay, rise time, and partition in a neural net analysis, and the last one exploits additional information from the fitted signal pulses to reconstruct recoil position and type. (10) The recoil energy threshold cut is set at 10 keV as a conservative effective discrimination of electromagnetic backgrounds. Most of the cuts were defined with $^{133}$Ba data and their efficiencies were calculated for neutrons from $^{252}$Cf data (or WIMP-search electron-recoil events for the ionization $\chi^{2}$ cut). The systematic uncertainties of the efficiencies were estimated by comparison of calibration with the WIMP-search electron recoils data. All the cut efficiencies were multiplied together to obtain the overall cut efficiency.

\section{Results of the 2004 campaign and current status of the experiment}

\begin{figure}[!h]
~\includegraphics[width=8.5cm]{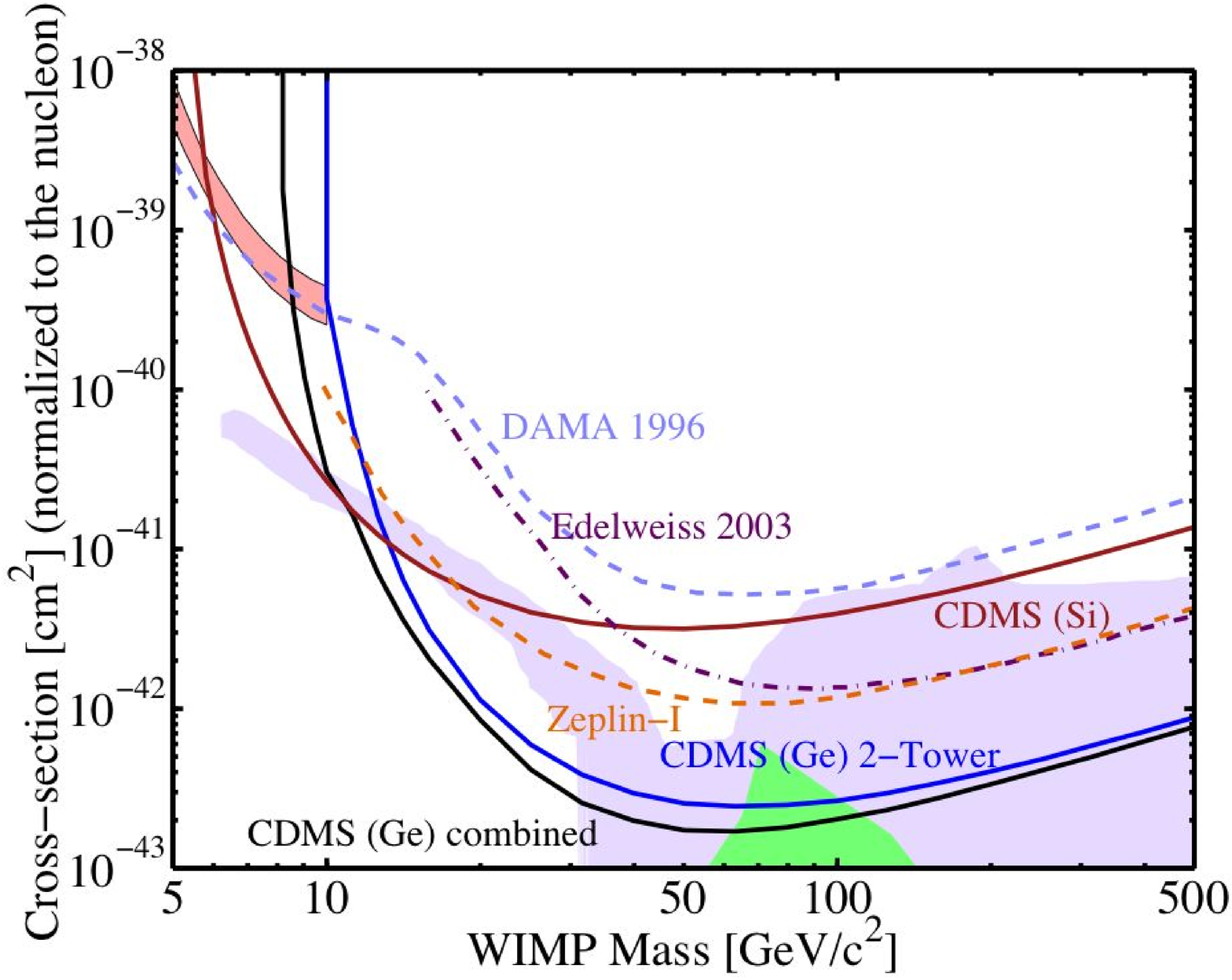}\includegraphics[width=7cm]{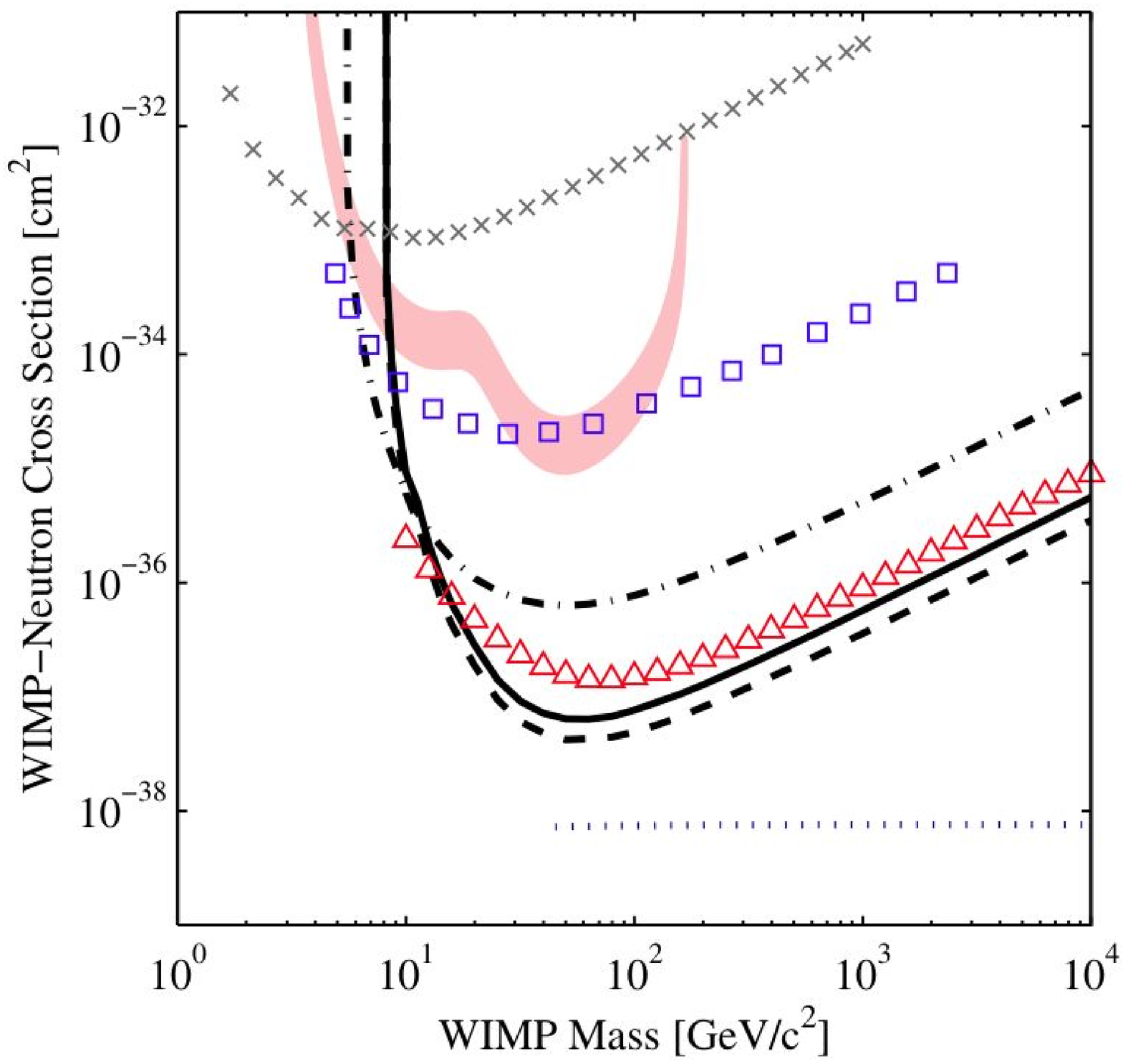}
\caption{\label{fig:sp}WIMP-nucleon cross-section upper limits (90\% C.L.) versus WIMP mass for the spin independent coupling (left) and the pure neutron coupling (right). On the left plot, the lowest curve uses Ge data from the combination of the two runs. MSSM~\protect\cite{bot} and CMSSM~\protect\cite{elis} models allow the shaded regions at the bottom. The shaded region in the upper left is from DAMA~\protect\cite{damar}, and experimental limits are from DAMA~\protect\cite{damal}, EDELWEISS~\protect\cite{edelsi} and ZEPLIN~\protect\cite{zepsi}. On the right plot, limits from the combined Soudan data are the solid line for Ge dash-dot lines for Si. Dashed curves represent Ge limits using an alternate form factor for the description of the nucleus. As benchmarks, we also included interpretations of the DAMA/NaI annual modulation signal~\protect\cite{damasd} (filled regions are 3{$\sigma$}-allowed) and limits from other leading experiments: CRESST~I~\protect\cite{cressd}~($\times$), PICASSO~\protect\cite{picassd}~({$\sqcap$}), ZEPLIN~I~\protect\cite{zepsd}~($\triangle$). EDELWEISS~\protect\cite{edelsd} and SIMPLE~\protect\cite{simplesd} limits are comparable to CDMS Si and PICASSO, respectively.}
\end{figure}

\begin{figure}[!b]
~~~\includegraphics[width=7cm]{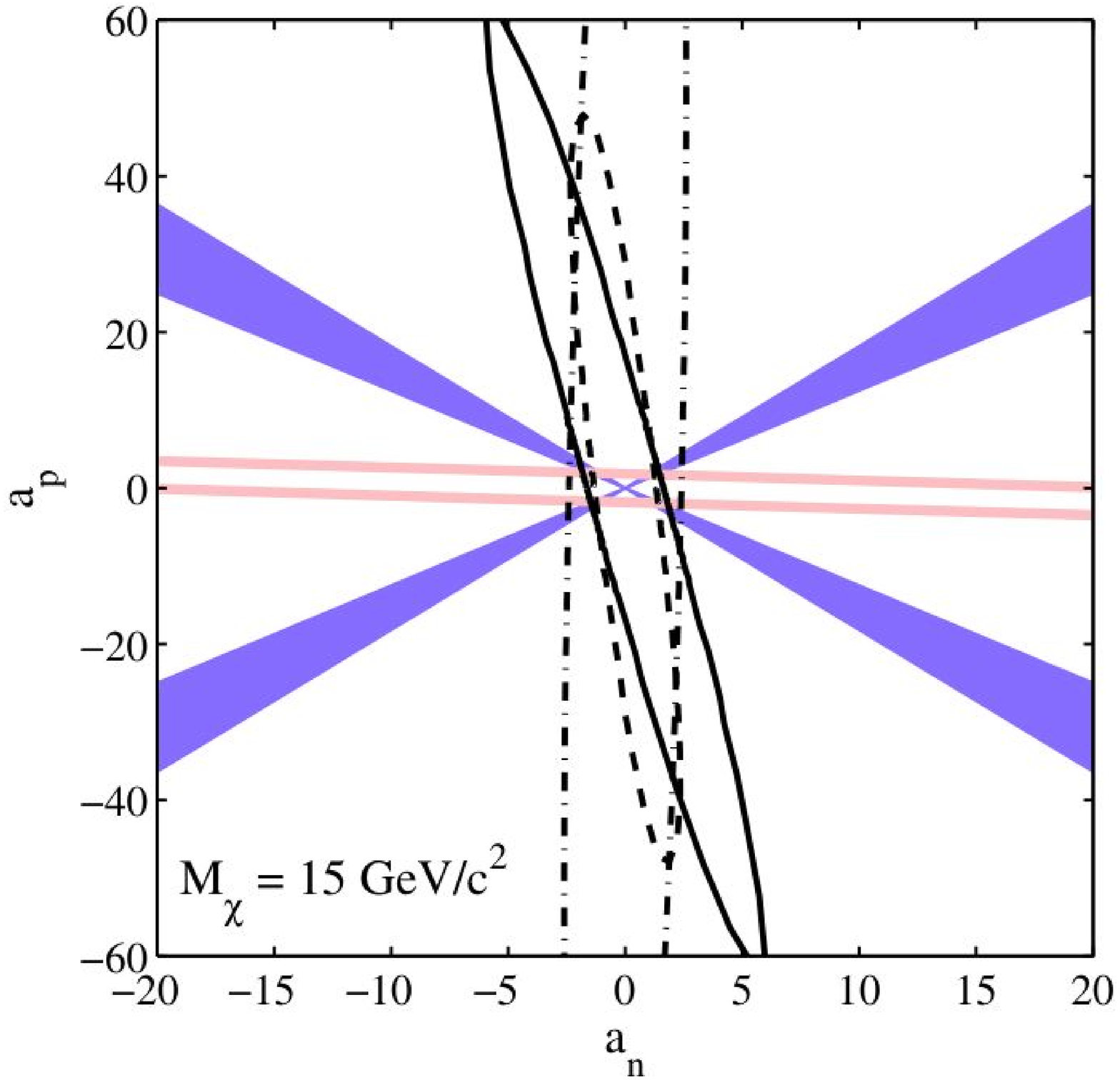}~~~~~~\includegraphics[width=7cm]{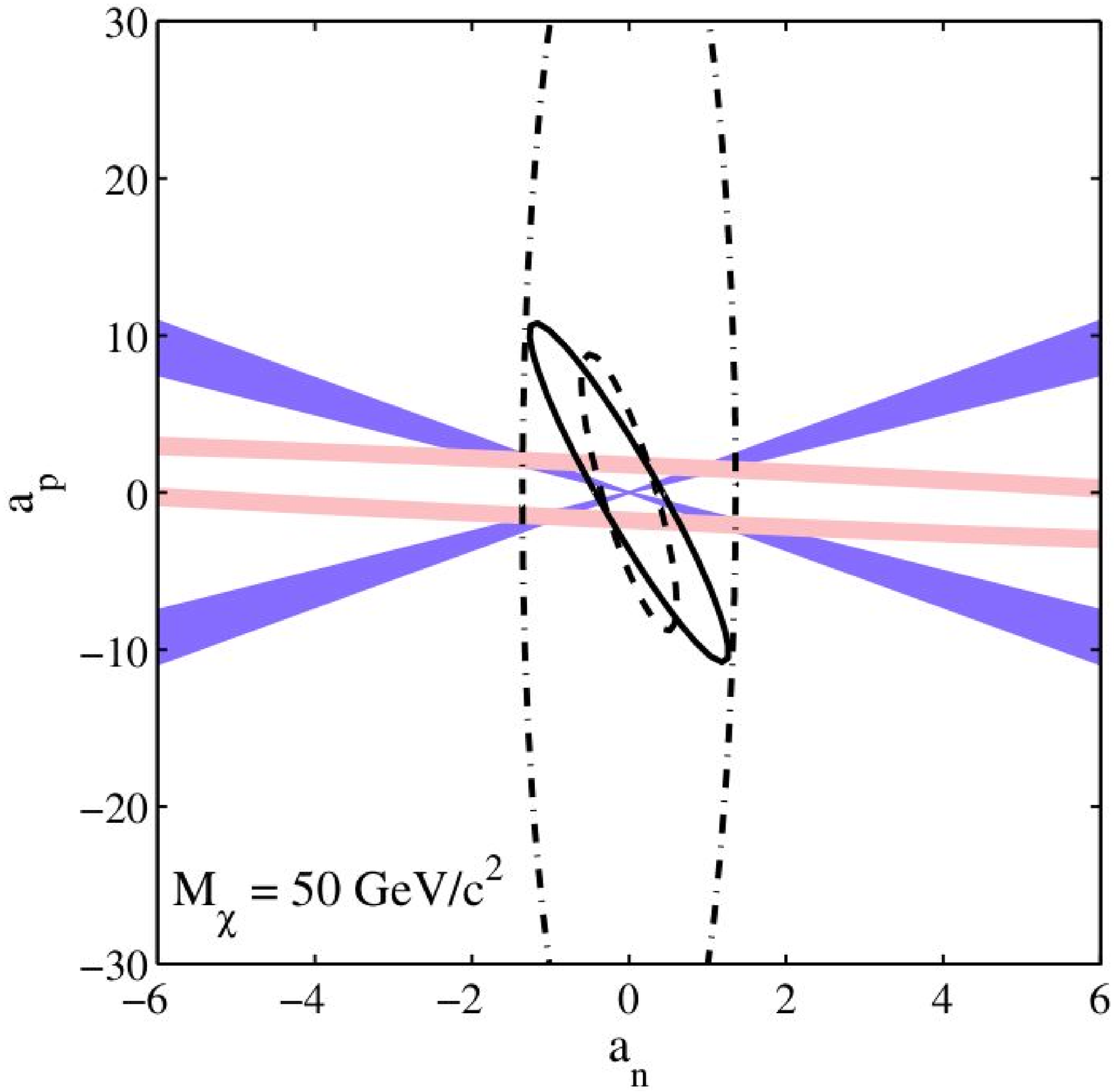}
\caption{\label{fig:anap}Regions in the $a_p-a_n$ plane allowed (90\% C.L.) by CDMS data. Each data set excludes the exterior of the corresponding ellipse. Two choices of WIMP mass are shown: 15 GeV/c$^{2}$ (left) and 50 GeV/c$^{2}$ (right). Dot-dashed ellipses represent Si limits, solid ellipses represent Ge limits, and dashed ellipses represent Ge limits using an alternate form factor. The near-horizontal light (pink) filled bands are the 3{$\sigma$}-allowed DAMA/NaI modulation signal. The thin dark (blue) filled wedges correspond to models satisfying $0.55 < {\vert}a_n/a_p\vert < 0.8$, a constraint from the effMSSM~\protect\cite{bed} framework.}
\end{figure}

\begin{figure}[!t]
~~~\includegraphics[width=7cm]{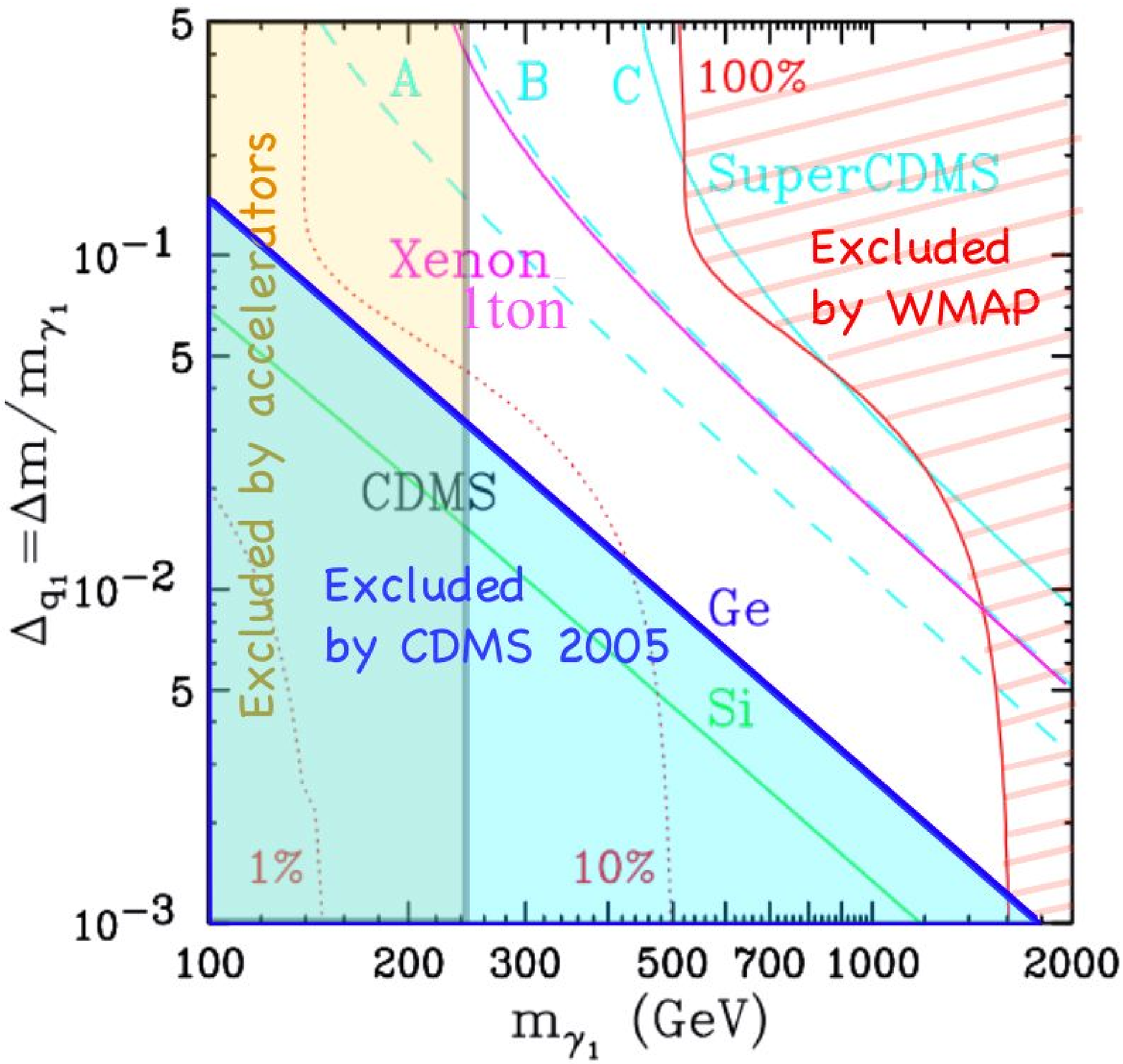}~~~~~~\includegraphics[width=7cm]{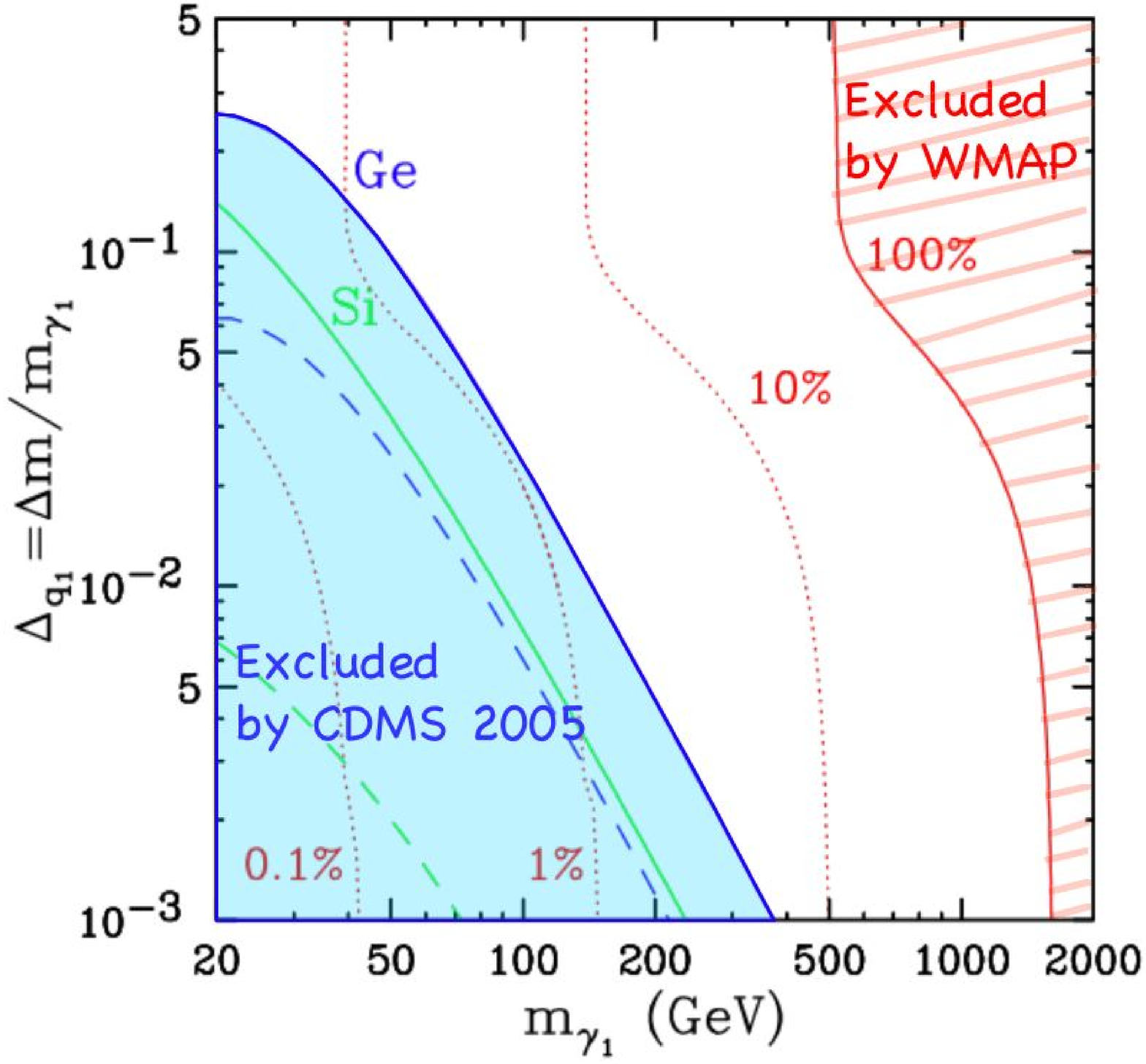}
\caption{\label{fig:ued}Experimental constrains on LPK in the LPK-quark mass ratios (${\Delta}_{q1}$) versus the LKP mass (${\gamma}1 \equiv B^{(1)}$) plane~\protect\cite{bkm}. The left plot shows the spin-independent region excluded in the last run by Ge (diagonal shaded region) and Si (diagonal line in the shaded region). The dashed region at the right of the curve is excluded by WMAP, the line labeled 100\% means that all WMAP allowed dark matter is made out of LKPs (${\Omega}_{LKP}$ = 0.27), while the 10\% and 1\% curves mean that only a fraction of the total dark matter density is in KK-particles. The right plot shows the same exclusion limits for the spin-dependent case.}
\end{figure}

We performed two runs at the Soudan mine. For both of them, the estimated number of neutrons that escaped the muon veto was less than 0.06 in Ge and in Si. Half of the {$\approx$}0.3 singles surface events per day observed were due to beta decays of contaminants and the other half to gamma rays. The first run, from October 2003 to January 2004, gathered a raw exposure of 52.6 kg{$\times$}d of WIMP-search data, with 4 Ge detectors and 2 Si detectors. One of the Si detectors was excluded from the analysis due to a $^{14}$C contamination. The exposure after analysis cuts was 19 kg{$\times$}days for the Ge detectors, in the 10-100 keV recoil energy band. One event passed all the cuts, consistent with the 0.7{$\pm$}0.3 expected misidentification rate of surface electron recoils~\cite{r118prl}$^{,}$~\cite{bgprd}. The second run, from March to August 2004, gathered 74.5 kg{$\times$}d of raw WIMP-search data, with 6 Ge detectors and 6 Si detectors. In addition to the contaminated Si detector, another Si and a Ge ZIPs, were excluded from the analysis because of poor phonon sensors performance. The exposure after analysis cuts was 34 kg{$\times$}days for the Ge detectors and 12 kg{$\times$}days for the Si detectors, in the 10-100 keV recoil energy band. One event passed all the cuts, consistent with the expected misidentification rate of surface electron recoils which was 0.4{$\pm$}0.2(stat){$\pm$}0.2(syst) for Ge and 1.2{$\pm$}0.6(stat){$\pm$}0.2(syst) for Si~\cite{siprl}$^{,}$~\cite{sdprd}. To report the Ge detectors results of this run we used the sum of delay and risetime surface event cut, and for the Si detectors the energy-dependent $\chi^{2}$ cut which had the best expected sensitivity to a nuclear recoil signal from low-mass WIMPs of any of our five methods. The other surface event timing analysis promised further improvements in sensitivity for the larger exposures planned in our future runs. For both runs the overall cut efficiency reached a plateau of $\approx$40\% above 20 keV for Ge, and $\approx$ 50\% above 40 keV for Si.\\
The upper limits on spin-independent WIMP-nucleon cross sections in the supersymmetric (SUSY) framework are shown in fig.~\ref{fig:sp}. They are calculated from the Ge and Si analyzes reported here, using standard assumptions for the galactic halo,  for the latest run and the combined runs. The combined result limits the WIMP-nucleon cross section to $<$1.6{$\times$}10-43 cm$^{2}$ at the 90\% C.L. at a WIMP mass of 60 GeV/c$^{2}$. This limit constrains some MSSM parameter space and for the first time excludes some parameter space relevant to constrained models (CMSSM).\\
Scaling the exposures by the isotopic abundances, we obtain a total of 11.5 raw kg{$\times$}d for $^{73}$Ge and 1.7 for $^{29}$Si for spin-dependent interactions~\cite{sdprd}. The upper-limit contours in the WIMP mass versus spin-dependent cross section plane in the cases of pure neutron coupling for the combined runs are shown in fig.~\ref{fig:sp}. Some regions of parameter space are excluded by no other experiment than CDMS in the case of pure WIMP-neutron coupling. To explore more general models, these results can also be expressed in the mixed coupling plane (WIMP-neutron $a_n$ vs WIMP-proton $a_p$) for various choices of WIMP mass. Two such choices (15 Ge/c$^{2}$ and 50 GeV/c$^{2}$) are shown in Fig.~\ref{fig:anap}. The upper limits set here do not yet constrain SUSY models.\\
In the UED framework, the best studied LPK candidate is $B^{(1)}$, the first KK-mode of the hypercharge gauge boson. The elastic scattering interaction between $B^{(1)}$ and nuclei in the extreme non-relativistic contains also spin-independent and spin-dependent terms. The upper limits on WIMP-nuclides cross-sections calculated in the UED framework from our Ge and Si analyzes are shown in fig.~\ref{fig:ued}. For the interpretation of our results the most notable assumptions~\cite{cfm}$^,$~\cite{st} are that the masses of the degenerated first level quarks, $q^{(1)}$, are taken as a free parameter, that the Higgs mass is 120 GeV/c$^{2}$, and that the ratio ${\Delta}_{q1} = (m_{q^{(1)}}-m_{B^{(1)}})/m_{B^{(1)}}$ is varied from 0.01 to 0.5. The CDMS results exclude the region to the left of the curve labeled "Ge" in the LKP-mass and ${\Delta}_{q1}$ parameter space, as seen in fig.~\ref{fig:ued}, and are complementary to accelerator limits and WMAP constraints.\\
In 2005 the experiment has been updated to operate 5 towers of detectors (19 Ge and 11 Si). After some technical delays, the experiment should be cooled again soon. Our estimate for the projected exposure after cuts based on two full years of acquisition is about 700 kg{$\times$}d Ge and 150 kg{$\times$}d Si. This should increase the cross-section limits by a factor {$\approx$}10.

\end{document}